\documentclass[reprint, superscriptaddress,
 amsmath,amssymb,
 aps, pra,
floatfix,linenumbers,]{revtex4-2}
\usepackage{physics} % using physics macros
\usepackage{amsmath} % math notation
\usepackage{graphicx}% Include figure files
\usepackage{dcolumn}% Align table columns on decimal point
\usepackage{bm}% bold math
\usepackage{lipsum}
\usepackage{xcolor, soul}

\usepackage{float}

\begin{document}
 \nolinenumbers 
\preprint{APS/123-QED}
\title{Programmable all optical spin simulator with artificial gauge fields}
\author{Simon Mahler}
\email{Email: sim.mahler@gmail.com} %% email address is required
\thanks{These authors contributed equally to this work.}
\author{Eran Bernstein}
\email{Email: eran.bernstein@gmail.com} %%
\thanks{These authors contributed equally to this work.}
\author{Sagie Gadasi}
\author{Geva Arwas}
\author{Asher A. Friesem}
\author{Nir Davidson}
\affiliation{Department of Physics of Complex Systems, Weizmann Institute of Science, Rehovot 761001, Israel}
\date{\today}% It is always \today, today, but any date may be explicitly specified

\begin{abstract}
The interconnection of lasers is pivotal across various research domains, from generating high-power lasers to studying out-of-equilibrium coupled systems. This paper explores our investigation into Hermitian coupling between lasers in an array, with the aim of achieving arbitrary coupling and creating artificial gauge fields that can break time-reversal symmetry. For that, we investigated Hermitian coupling within three laser array geometries: a square array of 100 lasers, a triangular array of 130 lasers, and a ring array of 8 lasers. In the square array, we implemented arbitrary laser coupling with a precision of $2\pi/120$ radians, enabling the attainment of any desired phase-locking state. In the triangular array, we controlled the chirality of the lasers with 99$\%$ purity. In the ring array, the introduction of an artificial gauge field revealed discrete quantized first-order transitions between distinct topological phase-locking states. This arbitrary coupling, with control over both the strength and phase, paves the way for exploring spin systems and configurations characterized by exotic, non-conventional coupling.\end{abstract}
\maketitle

Physical simulators are powerful tools for studying and understanding interacting many body Hamiltonians in the classical and quantum regimes~\cite{Nori2009,Guifre2004,Martin2021,Dario2020}. Spin simulators are particularly intriguing due to the universal aspects of spin Hamiltonians \cite{DeLasCuevas16,Cubitt18}, their special physical behavior ~\cite{Nixon13,Semeghini21}, and their mappings onto hard computational and optimization problems ~\cite{Lucas14,Martin2021,Dario2020,Yao2021, Berloff17, Berloff2022}. Indeed, they have been recently incorporated into a variety of physical systems, including ultracold atoms ~\cite{Zohar2016,Bloch2017,Schafer2020}, trapped ions~\cite{Bollinger2016,Blatt2012,Roos2011,Cirac2004,Kim2010} and super conducting circuits~\cite{Solano2014,Houck2012,Paraoanu2014}. Photonics is a most attractive platform for spin simulators as it offers accurate control, scalability, long range, dissipative and tunable coupling, room temperature operation and advanced fabrication tools. Some have already been exploited to develop large scale and well controlled optical spin simulators that are based on coupled lasers ~\cite{Pal20,Nixon13,Pierangeli19,Mahler19,Miri20,Arwas21}, optical parametric oscillators~\cite{Marandi14, McMahon16}, polariton Bose–Einstein condensates (BECs)~\cite{Berloff17,Berloff2022,Amo2015,Berloff20} and nano lasers~\cite{Parto2020_2,Parto2020}.

Most spin simulators do not have a real magnetic moment. Hence, to simulate their interactions with external magnetic fields, it is necessary to synthesize artificial gauge fields (AGFs) so as to obtain nonreciprocal systems, topological states, vorticity, flux quantization, fractional charges and additional exotic physical effects~\cite{Struck2013,Carusotto2011,Goldman2014,Jorg2020,Goldman2018,Lumer2019}. For spin simulators that are based on coupled optical oscillators, the AGFs correspond to Hermitian coupling between pairs of oscillators, namely the coupling $K$ between oscillators satisfy $K_{ij}=K_{ji}^{*}$ \cite{Carusotto2011, Mittal14}.

Some AGFs were implemented with coupled nano-photonic cavities and oscillators \cite{Lumer2019,Lustig2019,HAFEZI2014} where optical reciprocity was broken by non-Hermitian internal structure of individual oscillators and complex coupling was provided by waveguides with controlled lengths. In these, AGFs were clearly demonstrated, but the experimental systems had limited connectivity due to in-plane geometry, required high alignment tolerances, and had limited control of the coupling parameters. Pioneering works on time-multiplexed photonic resonator networks with controlled coupling were demonstrated in a long optical fiber system \cite{Leefmans22, Leefmans24} to study AGFs and topological lasing. However, these systems have not yet been utilized as spin simulators.

In this letter, we present a method for designing and generating AGF in an all-optical spin simulator that is based on an array of coupled lasers, formed in a compound degenerate ring cavity laser \cite{Pal20, Gershenzon20, Tradonsky21}. Degenerate cavity lasers exhibit nonlinear interactions between multiple intra-cavity lasing modes and optical gain, creating a system with numerous controllable parameters and a rapid reaction time \cite{Mahler20,Mahler21,Chriki18,Mahler20_2,Roadmap21,Chriki22}.
By introducing a proper phase-dependent local loss, we generate Hermitian coupling between adjacent lasers with a well controlled and digitally-programmable phase. We show below that the emerging minimal loss lasing solution, selected by mode competition \cite{Tradonsky19} corresponds to the ground state of an effective spin Hamiltonian with the desired AGF. 

We experimentally demonstrate the validity and efficacy of our method with three different laser array geometries: a square array for controlling the coupling in two-dimension, a triangular array for controlling the chirality of the lasers, and a ring array for controlling the topological charge of the lasers. For one, we use a square array of about a hundred lasers and dynamically scan the phase of the nearest neighbor coupling throughout the entire Brillouin zone with a precision of $2\pi/120$ radians. For the second, we use a triangular array of about 130 lasers and show how AGF can break the symmetry between staggered vortex and staggered anti-vortex states to obtain tunable chirality reaching $>$99$\%$ purity. Finally, we scan the flux of the AGF contained within a closed ring array of 8 lasers and observe discrete quantized flux plateaus connected by sharp transitions dictated by the periodic boundary conditions of the ring topology.

\begin{figure}[ht!]
\centering\includegraphics[width=0.4\textwidth]{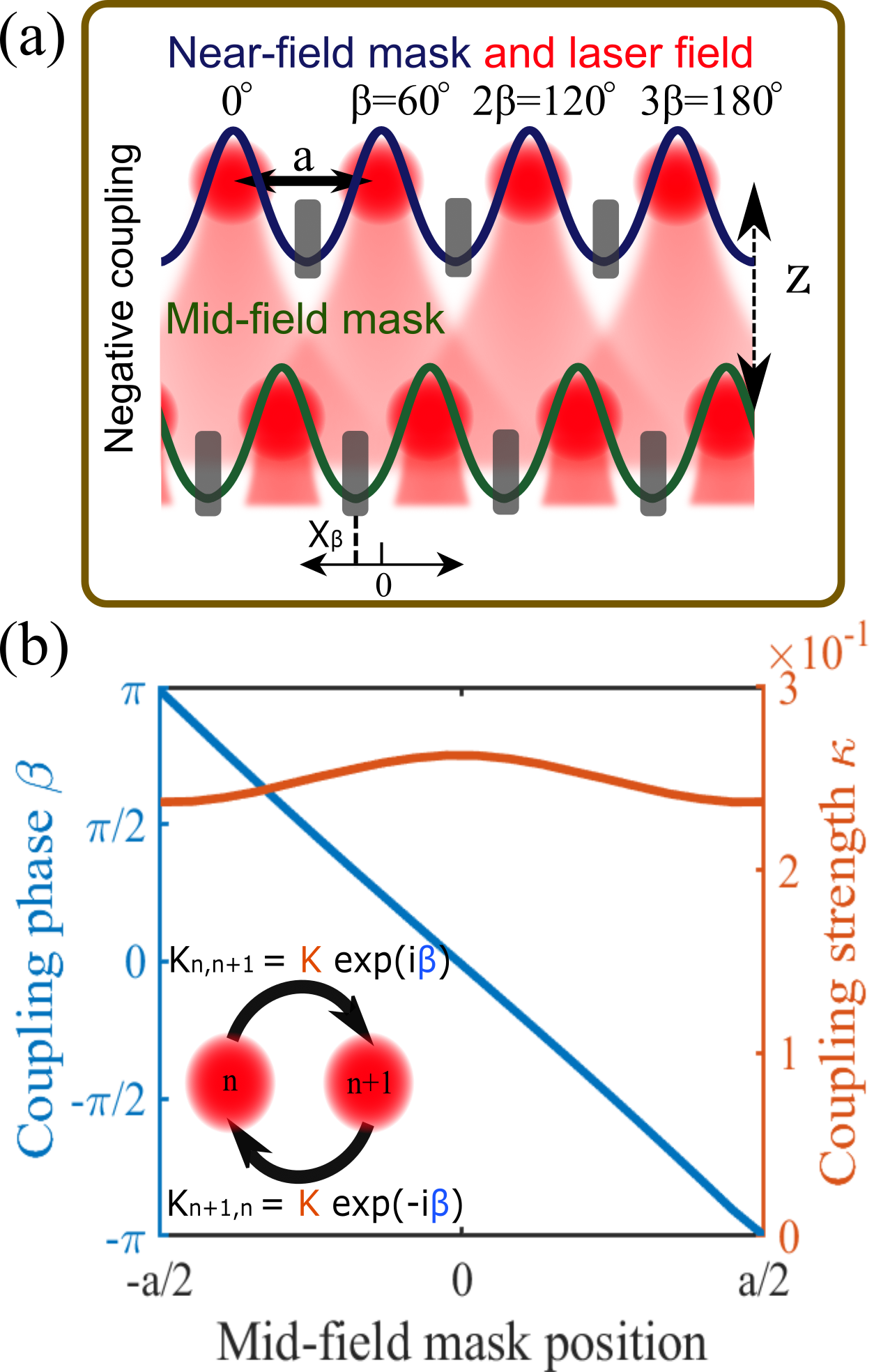}
\caption{Phase controlled coupling of lasers by intra-cavity mid-field absorptive mask. (a) Top: an intra-cavity near-field mask defines a linear array of 4 Gaussian lasers with separation $a$. Bottom: the mid-field absorptive mask, located $Z_{talbot}/2$ away from the near-field and shifted laterally by $X_{\beta}=a\frac{\beta}{2\pi}$ where it overlaps the dark fringes a the mid-field plane ensuring minimal loss lasing state. Here the phase difference $\beta$ between lasers is $\beta= 60^{\circ}$. (b) Numerically calculated coupling as a function of the mid-field mask position. The coupling strength (upper Gaussian curve) is symmetrical around the zero position of the mask, i.e. $\abs{K_{n, n+1}}=\abs{K_{n+1,n}}$, while the coupling phase (lower Linear curve) is Hermitian, i.e. $\beta_{n,n+1}=-\beta_{n+1,n}$, in agreement with the minimal loss picture described in (a).}
\label{fig:Fig_1_asymK}
\end{figure}

To understand the basic principle of our method, consider a 1D linear array of Gaussian-mode lasers with equal spacing $a$ between adjacent lasers in the "near-field"  (Fig. \ref{fig:Fig_1_asymK}(a)). After propagating half a Talbot distance $Z= Z_{Talbot}/2$, the distribution of each mode expands so as to significantly overlap the nearest neighbors (NN) laser fields. A phase difference $\beta$ between NN laser fields results in destructive interference and form dark fringes at a "mid-field" transverse position $X_{\beta}=a\frac{\beta}{2\pi}$ (see Fig. \ref{fig:Fig_1_asymK}(a)). Placing an absorptive mask at each of these locations with the highest absorption at $X_{\beta}$ ensures a minimal loss for a phase difference $\beta$ between NN lasers. With such a mid-field absorptive mask mode-competition \cite{Mahler22,Mahler21,Chriki18} ensures that the minimal-loss lasing solution has a phase-locked difference $\beta$ between NN, corresponding to a Hermitian coupling between NN, i.e. $K_{n,n+1}=\kappa e^{i\beta}$ and $K_{n+1,n}=\kappa e^{-i\beta}$ (Fig. \ref{fig:Fig_2_K_exp_arr}). See an analytic proof for arbitrary geometry in Supplementary Material~\cite{Supplemental} Section IV.

We numerically calculated the coupling \cite{Naresh22, Supplemental} $K_{n,n+1}=\kappa e^{i\beta}$ between NN lasers when using a binary mid-field mask with a fill factor of $=0.18$, while $X_{\beta}$ was varied from $-a/2$ to $a/2$ (Fig. ~\ref{fig:Fig_1_asymK}(b)). As evident, the calculated coupling strength is nearly constant while the anti-symmetric coupling phase $\beta_{n,n+1}=-\beta_{n+1,n}$ is exactly $2\pi \frac{X_{\beta}}{a}$, as predicted for the minimal loss lasing state. Increasing the fill factor of the mask and increasing the distance beyond $Z_{Talbot}/2$ enhances the coupling strength between NN but also increases the effects of NNN coupling (detailed in Supplementary Material \cite{Supplemental} Fig.~S1). This minimal-loss phase-locking method also applies to two-dimensional laser arrays where the Hermitian coupling corresponds to AGF.

\begin{figure}[ht!]
\centering\includegraphics[width=0.42\textwidth]{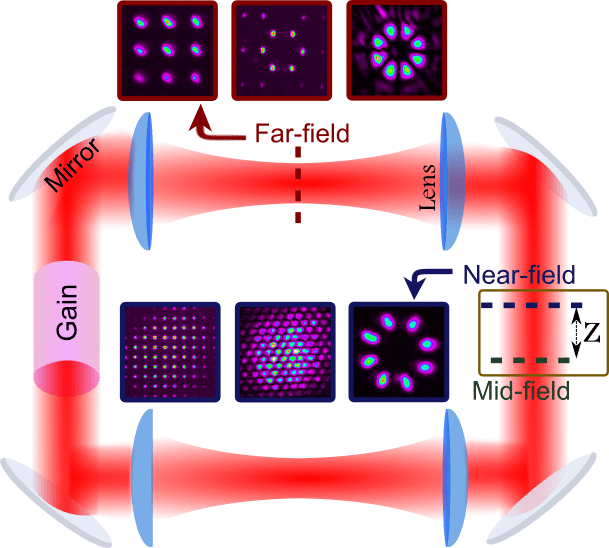}
\caption{DDRCL arrangement for implementing Hermitian coupling between NN lasers and for introducing an AGF. An intra-cavity near-field aperture mask defines the laser array geometry with their light propagating unidirectionally and self-imaged after a round trip by two 4f intra-cavity telescopes. A second aperture mask, located a distance $Z\sim Z_{Talbot}/2$ from the near-field mask, provides Hermitian coupling between NN lasers. Insets: experimental near-field intensity distributions of square, triangular and ring laser arrays and corresponding far-field intensity distributions. 
}
\label{fig:Fig_2_K_exp_arr}
\end{figure}

Our experimental arrangement for implementing Hermitian coupling between NN in large laser arrays is based on a digital degenerate ring cavity laser (DDRCL) ~\cite{Arnaud69,Tradonsky21,Davidson22}, schematically shown in Fig. ~\ref{fig:Fig_2_K_exp_arr} (see detailed arrangement in Supplementary Material \cite{Supplemental}, Fig.~S2). The light propagates unidirectionally, by resorting to an intra-cavity optical isolator (not shown in Fig. ~\ref{fig:Fig_2_K_exp_arr}) and is self-imaged after every round trip due to the two 4f intra-cavity telescopes. A first binary aperture mask located at the near-field plane defines an array of near-Gaussian lasers with any desired geometry e.g. square, triangular or ring arrays (see Fig.~\ref{fig:Fig_2_K_exp_arr} insets for near-field intensity distribution) \cite{Mahler19,Nixon13,Pal20}. A second aperture mask, located at the mid-field plane, a distance $Z\sim Z_{Talbot}/2=a^{2}/\lambda$ from the near-field plane, with lasing wavelength $\lambda=1064$ nm, provides the Hermitian coupling between NN lasers. In our experiments, the mid-field mask was implemented with an intra-cavity programmable spatial light modulator \cite{Tradonsky21,Mahler22}, so we could obtain precise digital control of either the laser array geometry or the coupling. 

\begin{figure}[ht!]
\centering\includegraphics[width=0.48\textwidth]{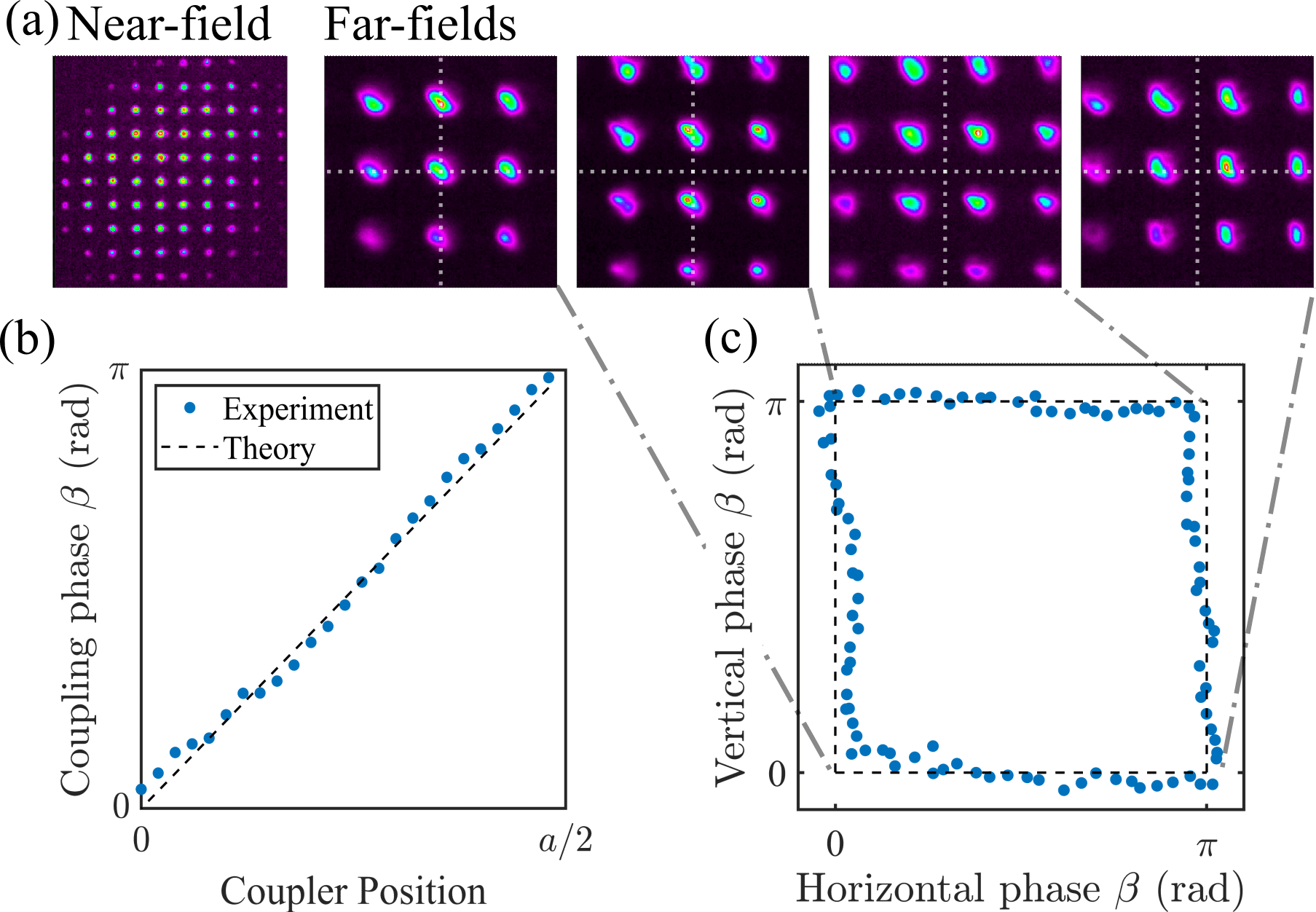}
\caption{Arbitrary phase locking of 100 lasers in a square array. 
(a) Experimental near-field and far-field intensity distributions of the array at four representative positions of the mid-field mask corresponding to horizontal and vertical phase differences between NN of $(0,0),(0,\pi), (\pi,\pi)$ and $(\pi,0)$, respectively.
(b) Experimental phase $\beta$ of the Hermitian coupling as a function of the coupling mask horizontal position $X_{\beta}$ showing excellent quantitative agreement with the theoretically predicted linear relation of Fig. \ref{fig:Fig_1_asymK}(b).
(c) Experimental horizontal and vertical coupling phases as the mid-field coupling mask is scanned along the edges of the first Brillouin zone (detailed in Supplementary Material \cite{Supplemental} Fig.~S3).
}
\label{fig:3_K_square}
\end{figure}

Using our DDRCL, we performed experiments with three different laser array geometries. First, we resorted to a square array of about 100 lasers where we varied the position of the intra-cavity square array mid-field absorptive mask (coupler), horizontally and/or vertically in order to control the phase of both the horizontal and vertical Hermitian coupling between NN lasers and force the entire array to phase lock. Representative results are shown in Fig. ~\ref{fig:3_K_square}. Figure \ref{fig:3_K_square}(a) shows the experimental near-field and far-field intensity distributions at four different positions of the mid-field coupler. The sharpness of the far-field diffraction peaks indicates good phase locking and their positions determine that the horizontal and vertical phase differences between NN lasers are $(0,0),(0,\pi), (\pi,\pi)$ and $(\pi,0)$. Figure ~\ref{fig:3_K_square}(b) shows the experimental phase of the Hermitian coupling as a function of the mask horizontal position $X_{\beta}$, indicating excellent quantitative agreement with the predicted linear relation $\beta_{n,n+1}=-\beta_{n+1,n}=2\pi \frac{X_{\beta}}{a}$, with RMS error of $ \approx\ 2\pi/120$  $(=\lambda/120)$. Fig.~\ref{fig:3_K_square}(c) presents the experimental phase of the Hermitian coupling when varied along a closed contour at the edges of the first Brillouin zone of the square array, demonstrating the ability to control arbitrary horizontal and vertical phases of the Hermitian coupling between NN lasers.

\begin{figure}[ht!]
\centering\includegraphics[width=0.4\textwidth]{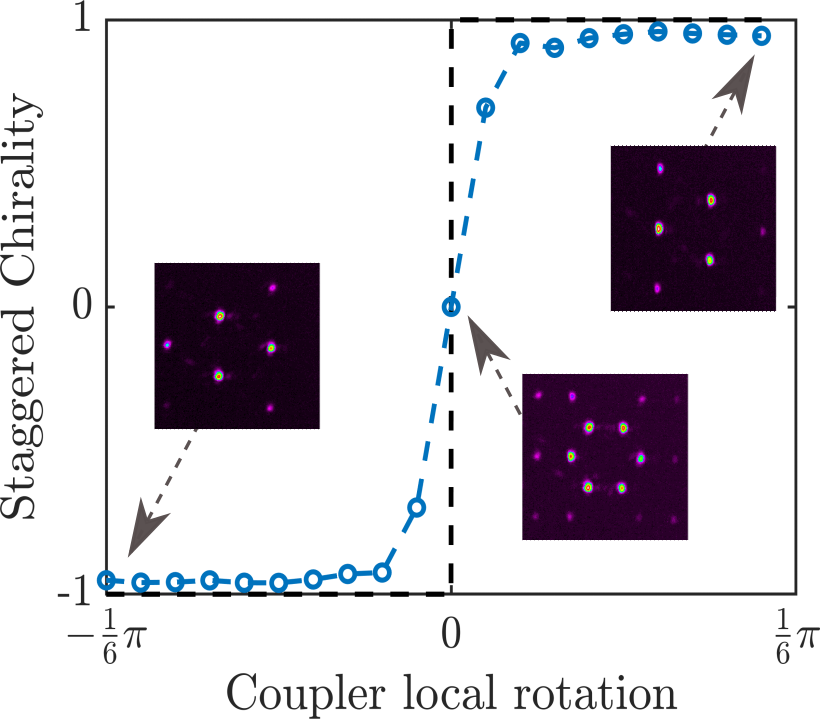}
\caption{AGF in a triangular lattice of 130 lasers (near-field intensity distribution shown in Fig. ~\ref{fig:Fig_2_K_exp_arr}, central inset). Experimental staggered chirality $C$  as a function of the coupler local rotation. Dashed black line: theoretical prediction for the chirality of the minimal loss lasing state.  Insets: experimental far-field intensity distributions. Left/right: 
near-pure staggered anti-vortex/vertex states, corresponding to near-pure negative/positive staggered chirality. Center: their equal superposition corresponding to near zero chirality.}
\label{fig:4_K_triangle}
\end{figure}

Next, we resorted to a triangular array of about 130 lasers where we combined a constant negative coupling between NN lasers using a far-field aperture and a controlled Hermitian coupling by locally rotating the apertures of the mid-field coupler, in order to get a phase-locked state with a controlled chirality. Each of the vortex/anti-vortex chiral states forms three sharp diffraction peaks in the far-field. For the symmetric negative coupling provided by the far-field aperture, these solutions are exactly degenerated yielding six near-equal far-field peaks with zero average chirality (central inset in Fig.~\ref {fig:4_K_triangle}). The Hermitian coupling phase $\beta$ is controlled by locally rotating the apertures of the mid-field coupler, with a triangular array symmetry, around their center by $\beta$ radians. The fractional Talbot distance of the mid-field coupler is $Z=Z_{Talbot}/4$. Fig. ~\ref{fig:4_K_triangle} shows the experimental staggered chirality $C=(P_{SV}-P_{SAV})/(P_{SV}+P_{SAV})$, where $P_{SV}$ and $P_{SAV}$ are the measured powers of the staggered vortex and staggered anti-vortex states respectively. The peaks in the far-field intensity distribution for the different phase-locked states are shown in Fig. ~\ref{fig:4_K_triangle} insets. The results show nearly pure negative chirality for negative coupler rotations ($\beta<0$) and nearly pure positive chirality for positive coupler rotations ($\beta>0$) with a sharp transition between them exactly at zero coupler rotation ($\beta=0$). A comparison between two mid-field masks, presented in Supplementary Material~\cite{Supplemental} Figs. S4 and S5 indicate very similar results.

\begin{figure}[ht]
\centering\includegraphics[width=0.44\textwidth]{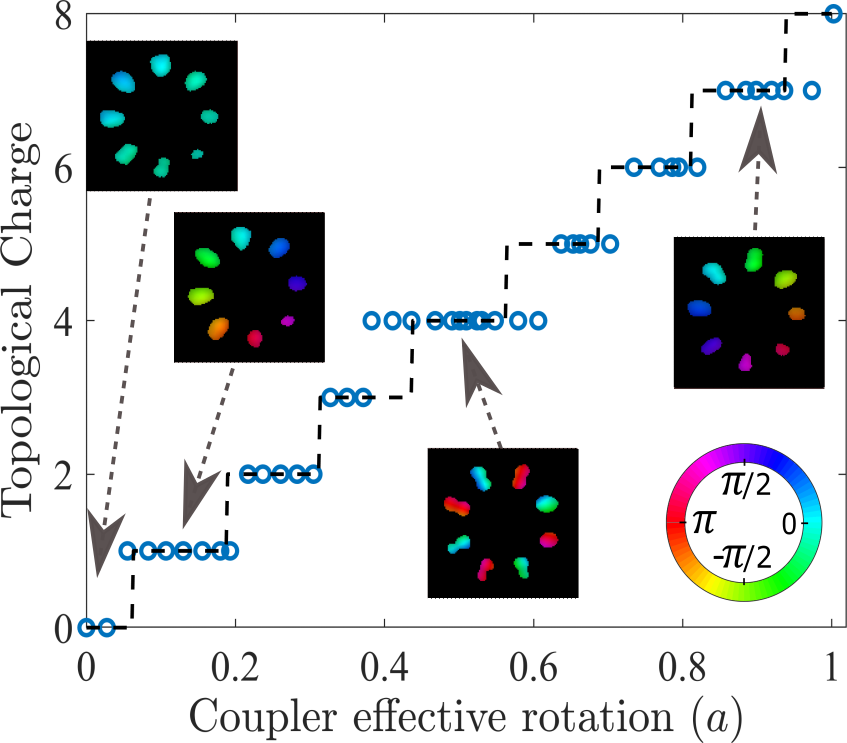}
\caption{Programmable artificial gauge field in a ring array of 8 lasers (near-field intensity distribution shown in Fig. ~\ref{fig:Fig_2_K_exp_arr}, right inset). Topological charge, obtained from the experimental near-field and far-field, as a function of the measured effective angle of rotation of the mid-field coupler relative to the laser array in the near-field. Dashed line: theoretical prediction for the topological charge of the minimal loss lasing state. Insets: reconstructed laser field at the near-field at four representative coupler rotations corresponding to topological charges of 0 (equivalent to 8), 1, 4 and 7 (equivalent to -1). Color represents the field reconstructed phase distribution obtained from the experimental near-field and the far-field intensity distributions using a procedure described in Supplementary Material \cite{Supplemental} section~V, and brightness represents the experimental field intensity distribution.}
\label{fig:5_Topo_ring}
\end{figure}

Finally, we control the Hermitian coupling between NN in a close ring array of 8 lasers by rotating a mid-field coupler, with the same geometry, at a distance $Z_{Talbot}/2$ such that its maximal absorption is located at $X_{\beta}=a\frac{\beta}{2\pi}$ as for the square array. We obtain the phase of all lasers by applying a Gerchberg-Saxton algorithm based on the experimental near-field and far-field intensity distributions and verify its validity by some direct interferometric measurements (see Supplementary Material ~\cite{Supplemental}, Figs.~S10 and ~S11).

The ring topology with its periodic boundary conditions restricts the possible phase-locked states to discrete vortex solutions with integer topological charge 
~\cite{Mahler19_2,Pal17} $Q=\frac{1}{2\pi}\sum_{n=1}^{N}\left\{\phi_{n+1}-\phi_{n}\right\}$, 
where $\{ \}$ wraps phases in the range from $\left[-\pi ~\mathrm{to} ~\pi\right]$ and $\phi_n$ is the phase of the laser number $n$. $Q=0$ corresponds to the in-phase locking state, $Q=\pm \text{int}(\frac{N}{2})$ corresponds to the out-of-phase locking state. And $2\pi \cdot Q$ is the quantized flux of the AGF through the ring.

Figure~\ref{fig:5_Topo_ring} shows the experimental topological charge of the ring laser array as a function of the mid-field coupler rotation. As expected, the topological charge reveals a series of sharp jumps between plateaus of integer values as the coupler angle is increased and the AGF flux reaches multiples of $2\pi$. Those results agree with the theoretical prediction marked in a dashed line. We believe that the differences between the obtained and the predicted topological charge are related to the internal structure of the mode inside each laser, as observed in Fig. \ref{fig:5_Topo_ring} insets showing the reconstructed laser phases. The discrete AGF flux and T.C. measured in the closed ring topology is analogous to the Harper-Hofstadter physics~\cite{Umucal11, Hofstadter76,Harper1955} and the integer quantum Hall effect~\cite{Ozawa19}.

To conclude, we introduced and demonstrated an approach for scalable and tunable Hermitian nearest neighbor coupling in large laser arrays, acting as all-optical spin simulators~\cite{Pal20, Gershenzon20}, using a digitally programmable intra-cavity absorption mask ~\cite{Davidson22,Tradonsky21}. 
Remarkably, our paraxial out-of-plane coupling configuration provides $\sim$100 fold reduction of the sensitivity of our coupling phase on the lateral mask position as compared to in-plane coupling schemes ~\cite{HAFEZI2014}. This inherent robustness enables extremely accurate control of the coupling phase, enabling us to demonstrate $2\pi/120$ accuracy even in large laser arrays. 
We showed how this highly accurate and tunable Hermitian coupling can produce an AGF that can break the degeneracy between staggered vortex and anti-vortex states and provide $99\%$ pure chiral states in large triangular arrays. Finally, we showed that in a close ring geometry with periodic boundary conditions, the applied AGF generates robust discrete vortex states with programmable integer topological charge and flux. 

Our photonic spin simulators with programmable and highly accurate AGF, combined with intra-cavity adaptive optics \cite{Tradonsky21,Davidson22}, tunable disorder \cite{Mahler22,Mahler21} and on-site complex potentials \cite{Arwas21,Pal20} can be used to study a plethora of topological, non-Hermitian and non-reciprocal phenomena, as well as provide large coherent laser arrays for photonics computational applications and high brightness laser sources. Our method can be readily extended to form Hermitian coupling non uniform strength or phase by using non homogeneous mid-field masks, and even non-Hermitian arbitrary coupling with mid-field phase masks.

\begin{acknowledgments}
The authors thank Chene Tradonsky and Sasha Poddubny for valuable help. The authors also thank Israel Science Foundation (1520/22), Minerva foundation, and the joint NSFC-ISF Foundation (3652/21).
\end{acknowledgments}

\bibliographystyle{apsrev4-2}
\bibliography{bibliography.bib} % Produces the bibliography via BibTeX.

\begin{thebibliography}{65}
\bibitem [{Sup()}]{Supplemental}%
  \BibitemOpen
  \bibinfo {note} {{S}ee {S}upplemental {M}aterial at [URL will be inserted by
  publisher] for further technical details, numerical simulations, and
  experimental results}\BibitemShut {NoStop}%
\end{thebibliography}
\end{document}